\documentclass[twocolumn]{aastex631}

\usepackage{graphicx}   
\usepackage{dcolumn}    
\usepackage{bm}         
\usepackage{verbatim}   
\usepackage{booktabs}
\usepackage{amsmath,amssymb}
\usepackage{mathrsfs}
\usepackage{amsfonts}
\usepackage{color}
\usepackage{natbib}
\usepackage{textcomp}
\usepackage{etoolbox}
\usepackage{floatrow}
\usepackage{multirow}
\usepackage{inputenc}
\usepackage{CJK}

\bibpunct{(}{)}{;}{a}{}{,}
\usepackage{floatrow}
\usepackage[caption=false]{subfig}
\usepackage{savesym}
\savesymbol{tablenum}
\usepackage{siunitx}
\restoresymbol{SIX}{tablenum}
\usepackage{enumitem}   

\usepackage[normalem]{ulem}

\shorttitle{Conservation of Total Wave Action in the Expanding Solar wind}
\shortauthors{Z.-S. Huang et al.}

\usepackage{subfiles}
\usepackage{xr} 
\graphicspath{{./}{figures/}}

\begin{document}
\begin{CJK*}{UTF8}{gbsn}

\title{Conservation of Total Wave Action in the Expanding Solar Wind}

\correspondingauthor{Zesen Huang}
\email{zesenhuang@g.ucla.edu}

\author[0000-0001-9570-5975]{Zesen Huang (黄泽森)}
\affiliation{Earth, Planetary, and Space Sciences, \;
University of California, Los Angeles \\
Los Angeles, CA 90095, USA }

\author[0000-0002-2582-7085]{Chen Shi (时辰)}
\affiliation{Earth, Planetary, and Space Sciences, \;
University of California, Los Angeles \\
Los Angeles, CA 90095, USA }

\author[0000-0002-1128-9685]{Nikos Sioulas}
\affiliation{Earth, Planetary, and Space Sciences, \;
University of California, Los Angeles \\
Los Angeles, CA 90095, USA }

\author[0000-0002-2381-3106]{Marco Velli}
\affiliation{Earth, Planetary, and Space Sciences, \;
University of California, Los Angeles \\
Los Angeles, CA 90095, USA }



\begin{abstract}
The conservation of wave action in moving plasmas has been well-known for over half a century. However, wave action is not conserved when multiple wave modes propagate and coexist close to degeneration condition (Sound speed equals Alfv\'en speed, i.e. plasma $\beta \sim 1$). Here we show that the violation of conservation is due to wave mode conversion, and that the total wave action summed over interacting modes is still conserved. Though the result is general, we focus on MHD waves and identify three distinctive mode conversion mechanisms, i.e. degeneracy, linear mode conversion, and resonance, and provide an intuitive physical picture for the mode conversion processes. We use 1D MHD simulations with the Expanding Box Model to simulate the nonlinear evolution of monochromatic MHD waves in the expanding solar wind. Simulation results validate the theory; total wave action therefore remains an interesting diagnostic for studies of waves and turbulence in the solar wind.
\end{abstract}



\section{Introduction}\label{sec:intro}

The heliosphere is permeated by the solar wind, a supersonic and super-Alfv\'enic plasma flow originating from the solar corona, and continuously expands into the interplanetary medium \citep{parker_dynamics_1958,velli_supersonic_1994}. Since the beginning of {\it in situ} observations, it has been confirmed by various studies \citep{coleman_wave-like_1967,coleman_turbulence_1968,belcher_alfvenic_1971,belcher_large-amplitude_1971} that the interplanetary space is filled with Alfv\'enic MHD turbulence and compressive fluctuations like the Pressure Balanced Structures (PBS) \citep{marsch_mhd_1991,tu_mhd_1995}. Over the years, numerous studies have been conducted on the Alfv\'enic fluctuations in the solar wind, showing that interplanetary Alfv\'en waves are "Arc Polarized" or "Spherically Polarized" \citep{tsurutani_relationship_1994,riley_alfvenic_1995,tsurutani_nonlinear_1997,bale_highly_2019,tenerani_evolution_2021}, kinetic in Nature \citep{tsurutani_review_2018}, and exhibit rich nonlinear effects \citep{hollweg_density_1971,tsurutani_review_2018,stefani_mode_2021}. On the other hand, magnetosonic waves are more scarce, with some exceptions including in at the upstream of interplanetary shocks \citep{tsurutani_waves_1983}, which are likely generated locally by the instabilities associated with upstream beams of energetic ions; proton cyclotron waves generated locally by the kinetic dissipation of the nonlinear Alfv\'en wave \citep{tsurutani_phase-steepened_2002}, which in the low frequency limit becomes slow magnetosonic waves; and in the solar corona [see e.g. \cite{ofman_slow_1999,pascoe_fast_2013,yang_numerical_2015}] Note however, with the plane-wave assumption, the fluctuations in the solar wind have non-negligible magnetosonic waves composition \citep{chaston_mhd_2020,zhu_wave_2020}. Therefore, the nonlinear evolution of magnetosonic waves in the solar wind remains an interesting topic.

Basic to the understanding of the wave evolution in the highly structured solar wind is the comprehension of the simpler, isotropic case, i.e., that of evolution in a plain, isotropic radial expanding wind. This obviously simple problem is not well-known yet. In the linear case, only the evolution of Alfv\'en waves is well understood: the Wentzel--Kramers--Brillouin (WKB) approximation predicts a 1/R decrease of the specific energy \citep{whang_alfven_1973}. However, the WKB approximation (as well as the finite frequency approximations, \citep{heinemann_non-wkb_1980,velli_waves_1991,velli_propagation_1993}, are not able to cope with the mode mixing introduced by the expansion \citep{lou_propagation_1993,lou_three-dimensional_1993,lou_three-dimensional_1993-1}. The coupling arises because (a) The characteristics of different degrees of freedom (Alfv\'enic, Slow, Fast) depends on the plasma $\beta=\frac{2\mu_0 p}{B^2}$ which changes with distance; (b) The d.c. (background) magnetic field $\boldsymbol B_0$ and wave vector $\boldsymbol k$ change both in direction and modulus due to the expansion, which further modifies the MHD eigenmodes polarization; (c) Different modes tend to decay differently with the expansion, and so does higher degree effects such as wave steepening, and relative strength of wave-coupling.

Moreover, for an infinitely long monochromatic MHD wave train propagating in expanding medium, another underknown effect further complicates the situation. Contrary to common knowledge, the adiabatic invariant of the wave train (Wave Action) \citep{whitham_general_1965,f_p_bretherton_wavetrains_1968,dewar_interaction_1970} is not well-conserved if the background conditions evolve close to degeneration point (Alfv\'en speed $v_a$, Sound speed $c_s$, wave vector $\vec k$, and background magnetic field $\vec B_0$ simultaneously satisfy: $v_a = c_s$ and $\vec k \parallel \vec B_0$) even in the WKB limit. This special condition can be easily achieved if the medium expands, e.g. in the expanding solar wind (see Figure {\ref{fig:EBM}}) where the plasma $\beta \sim 1$. This topic has not been covered thoroughly in past literature, especially for magnetosonic modes, partially because of their dissipative nature. Early studies \citep{jacques_momentum_1977,lou_three-dimensional_1993} on this subject mainly focused on their WKB evolution, i.e. {\it a priori} assumption of wave action conservation. Some other studies focused more on predicting the magnetogravity mode-conversion rate \citep{zhugzhda_magnetogravity_1979, zhugzhda_conversion_1981, zhugzhda_transformation_1982,zhugzhda_linear_1982,cally_note_2001, mcdougall_mhd_2007,mcdougall_new_2007,mcdougall_mhd_2009}. On the other hand, the subject of wave action conservation itself is more of theoretical interest and has only been studied in a general sense by \citep{hirota_wave-action_2010}. Therefore, a thorough study of the evolution of simple MHD waves in expanding solar wind is still lacking. Our study aims to provide an intuitive physical picture of the mechanisms behind the violation of conservation law for infinitely long monochromatic wave train. 

In this study, we propose a simple model to address the violation of wave action conservation. Our model shows that the violation is due to wave mode conversion, and that the total of wave action summed over all interacting modes (Alfv\'en, Slow, Fast) is a universally conserved quantity. In addition, we propose three distinctive mechanisms of the mode conversion, i.e. degeneracy, linear mode conversion, and resonance, providing an intuitive physical picture explaining the mode conversion process. By generalizing the conservation law for wave action, our model can serve as an extension of classical wave action conservation theory.

The rest of this paper is organized as follows: In section 2, we start by reviewing the theory for the conservation of wave action in MHD and propose a simple, intuitive model for wave mode conversion and conservation of total wave action; in section 3, we present complementing simulation results to substantiate our model; in section 4, we discuss the bifurcated behaviours of Alfv\'en mode and magnetosonic modes; in section 5, we summarize our results.

\section{Theory}\label{sec:theory}

In this section we give a brief overview of the concept of wave action \citep{whitham_general_1965,f_p_bretherton_wavetrains_1968,dewar_interaction_1970} with MHD equations, and suggest a possible scenario leads to violation of wave action conservation. And we propose a simple, intuitive model showing that the total of wave action summed over all interacting modes is a universally conserved quantity.

\subsection{Wave Action}
The Lagrangian density for MHD system is \citep{lundgren_hamiltons_1963}:
\begin{eqnarray}
    \begin{aligned}
        \mathcal{L}&=\frac{1}{2}\rho U^2-\frac{p}{\gamma-1}-\frac{B^2}{2\mu_0}
    \end{aligned}
    \label{Lagrangian_Density}
\end{eqnarray}
Where $\rho, \vec U, p, \vec B, \gamma$ are density, flow velocity, pressure, magnetic field, and adiabatic gas constant. To study the perturbation behaviors of this system, we decompose all fields into the background part plus the perturbation part. In this study, we limit the perturbations to be small compared with background fields. We adopt a WKB style temporal scale separation (wave frequency within the MHD regime but much higher than the effective frequency of expansion time scale). First, expand the Lagrangian density ($\mathcal{L}=\mathcal{L}_0+\mathcal{L}_1+\mathcal{L}_2+o(\delta^2)$); Second, discard the first-order terms because they average to zero (both temporally and spatially); Last, keep the second-order terms [for details, see \cite{dewar_interaction_1970}]:
\begin{eqnarray}
    L = \mathcal{L}_2 = \frac{1}{2}  \rho_0 (\Delta \vec u)^2-\frac{1}{2}\frac{(\Delta p)^2}{c_s^2\rho_0}-\frac{(\Delta\vec B)^2}{2\mu_0}
    \label{L2}
\end{eqnarray}
where quantities with subscript "0" are the background fields, and quantities with $\Delta$ are the perturbations ($\Delta f=f-f_0$, and $f_0=\langle f \rangle$). $c_s=\sqrt{\gamma p_0/\rho_0}$ is the sound speed. To proceed, we need to substitute all perturbations with their Fourier-transformed counterpart. The full ideal-MHD equation set with adiabatic closure is:
\begin{eqnarray}
        \frac{\partial \rho}{\partial t}+\nabla\cdot(\rho \boldsymbol u)=0 
        \label{eqn:conservation of mass}\\
        \rho\left(\frac{\partial \boldsymbol u}{\partial t}+\boldsymbol u \cdot \nabla \boldsymbol u \right)=-\nabla p+ \frac{1}{\mu_0}(\nabla \times \boldsymbol B)\times \boldsymbol B\\
        \frac{\partial \boldsymbol B}{\partial t}=\nabla\times(\boldsymbol u\times \boldsymbol B)
        \label{eqn:faraday's law}\\
        \nabla\cdot\boldsymbol B=0\\
        \frac{d}{dt}\left(p\rho^{-\gamma}\right)=0
        \label{eqn:adiabatic closure}
\end{eqnarray}
The displacements of three MHD eigenmodes form an orthogonal triad, and hence without loss of generality, we write the flow perturbation of mode $M$ as:
\begin{eqnarray}
    \Delta \vec u_M = \tilde{a}_M \omega_M \hat e_M
    \label{eqn:displacement}
\end{eqnarray}
After linearization, plug (\ref{eqn:displacement}) into (\ref{eqn:conservation of mass}) and (\ref{eqn:adiabatic closure}), we obtain:
\begin{eqnarray}
    \Delta p_{M} = c_s^2\delta \rho_M =  \tilde{a}_M c_s^2 \rho_0 k_M(\hat k_M \cdot \hat e_M)
    \label{eqn:deltap}
\end{eqnarray}
and into (\ref{eqn:faraday's law}), we obtain:
\begin{eqnarray}
    \Delta \vec B_{M} = \tilde{a}_M B_0 k_M\left[\hat b (\hat k_M \cdot \hat e_M)-(\hat b \cdot \hat k_M)\hat e_M\right]
    \label{eqn:deltaB}
\end{eqnarray}
where $\tilde{a}_M$ is complex amplitude of displacement, $\omega_M$ is intrinsic frequency of the wave, $\vec k_M$ is wave vector, $\hat e_M$ is the unit vector along displacement, and $\hat k_M = \vec k_M/k_M$, $\hat b = \vec B_0/B_0$ are unit vectors of wave vector and background magnetic field, all of mode $M$.

Finally we plug (\ref{eqn:displacement})-(\ref{eqn:deltaB}) into (\ref{L2}) and temporally or spatially average it and obtain the averaged Lagrangian Density $\mathscr{L}$:
\begin{eqnarray}
    \begin{aligned}
        \mathscr{L}_M(\tilde a_M, &-\partial_t \theta_M, \nabla_x \theta_M)
        \\
        =&\frac{1}{4}\rho_0 \tilde a_M^2\left\{
        \omega_M^2-c_s^2 k_M^2 (\hat k_M \cdot \hat e_M)^2\right.
        \\
        &\left. -v_a^2 k_M^2\left[\hat b (\hat k_M \cdot \hat e_M)-(\hat b \cdot \hat k_M)\hat e_M\right]^2\right\}
    \end{aligned}
    \label{averaged_lagrangian}
\end{eqnarray}
where $\theta_M(x,t)$ is the wave phase, hence $-\partial_t\theta_M=\omega_M$ and $\nabla_x\theta_M = \vec k_M$. Note that for Alfv\'en mode ($\delta \rho_A=0$, $\vec k_A \cdot \hat e_A=0$),  the Lagrangian density can be reduced to:
\begin{eqnarray}
    \begin{aligned}
        \mathscr{L}_A(\tilde a_A, &-\partial_t \theta_A, \nabla_x \theta_A)
        \\
        &=\frac{1}{4}\rho_0 \tilde a_A^2\left[\omega_A^2- v_a^2 k_A^2 (\hat b \cdot \hat k)^2\right]
    \end{aligned}
\end{eqnarray}

\citep{whitham_general_1965,f_p_bretherton_wavetrains_1968} have shown that for a slowly varying (WKB) wavetrain, the local amplitude, frequency, and wavenumber are governed by the variational principle (henceforward we change the notations: $\partial_t \theta\to \theta_t$ and $\nabla_x \theta\to \theta_x$):
\begin{eqnarray}
    \delta \int \mathscr{L}(\tilde a, -\theta_t, \theta_x) dx dt = 0
    \label{eqn:whitham variantion principle}
\end{eqnarray}
subject to infinitesimal variations $\delta \tilde a(x,t)$, $\delta \theta(x,t)$ which vanish at infinity. Variation with respective to $\tilde a$ yields ($\mathscr{L}=\tilde a^2 \bar{\mathscr{L}}$):
\begin{eqnarray}
    \begin{aligned}
    &\frac{\partial \mathscr{L}}{\partial \tilde a}=2\tilde a \bar{\mathscr{L}}=0\\
    & \Rightarrow \mathscr{L}=0
    \end{aligned}
\end{eqnarray}
which is equivalent to the dispersion relations. Variation with respect to $\theta$ on (\ref{eqn:whitham variantion principle}) yields (see Appendix-A for detailed derivation):
\begin{eqnarray}
    \begin{aligned}
        \frac{\partial}{\partial t}\left(\frac{\partial \mathscr{L}}{\partial \omega}\right)-\frac{\partial}{\partial x}\left(\frac{\partial \mathscr{L}}{\partial k}\right)=0
    \end{aligned}
\end{eqnarray}
This is a conservation equation for the quantity $\partial \mathscr{L}/{\partial \omega}$ subject to flux $-\partial \mathscr{L}/\partial k$. 
Now substitute $\theta$ with $\theta_{M}$, $k$ with $\vec k_M$, and rewrite $\partial_x$ as $\nabla$, we have:
\begin{eqnarray}
\frac{\partial}{\partial t}\left(
\frac{\partial \mathscr{L}_{M}}{\partial \omega_{M}}\right)
-\nabla \cdot\left(
\frac{\partial \mathscr{L}_{M}}{\partial \vec k_{M}}
\right)=0
\end{eqnarray}
Considered that the dispersion relations are equivalent to:
\begin{eqnarray}
\mathscr{L}_{M}=0
\end{eqnarray}
and the group velocities are:
\begin{eqnarray}
\vec v_{g,M}=-\left.\frac{\partial \mathscr{L}_{M}}{\partial \vec k_{M}}\right/\frac{\partial \mathscr{L}_{M}}{\partial \omega_{M}}
=
\mathscr{L}_{\vec k, {M}}/\mathscr{L}_{\omega, {M}}
\end{eqnarray}
So that the conservation equation turns into:
\begin{eqnarray}
\frac{\partial}{\partial t}\left(
\frac{\partial \mathscr{L}_{M}}{\partial \omega_{M}}\right)
+\nabla \cdot\left(
\vec v_{g,M}
\frac{\partial \mathscr{L}_{M}}{\partial \omega_{M}}
\right)=0
\label{conservation}
\end{eqnarray}
(\ref{conservation}) marks the conservation law for wave action density $\mathscr{L}_{\omega, {M}}$, subject to flux $\vec v_{g,M} \mathscr{L}_{\vec k, {M}}$. The wave energy density can be further defined as:
\begin{eqnarray}
    \begin{aligned}
        \mathscr{E}&=\omega \mathscr{L}_\omega-\mathscr{L}\\
        &=\frac{1}{2}  \rho_0 \langle(\Delta \vec u)^2\rangle+\frac{1}{2}\frac{\langle(\Delta p)^2\rangle}{c_s^2\rho_0}+\frac{\langle(\Delta\vec B)^2\rangle}{2\mu_0}
    \end{aligned}
\end{eqnarray}
and consider that for waves with small amplitude $\mathscr{L}=0$, the wave action density $h_M$ for mode $M$ is defined as:
\begin{eqnarray}
h_M = \mathscr{L}_{\omega,{M}} = \frac{\mathscr{E}_{M}}{\omega_{M}}
\end{eqnarray}
where $\mathscr{E}_{M}$ is the wave energy density and $\omega_{M}$ is the intrinsic frequency of Alfv\'en, Slow, and Fast wave respectively. And finally we have the {\it conservation of wave action for monochromatic waves}:
\begin{eqnarray}
    \frac{\partial}{\partial t}\left(
    \frac{\mathscr{E}_{M}}{\omega_{M}}\right)
    +\nabla \cdot\left(
    \vec c_{M}
    \frac{\mathscr{E}_{M}}{\omega_{M}}
    \right)=0    
\end{eqnarray}
Integrating in space and assuming periodicity at the boundary, we get:
\begin{eqnarray}
    \hbar_M=\frac{E_M}{\omega_M}=const.
\end{eqnarray}
where $E_M=\int_V \mathscr{E}_M d\nu$ and $\hbar_M$ is the wave action (quantum) for mode $M$. Note that wave action is the counterpart of adiabatic invariant for waves in fluid system and is independent of the detailed description (e.g. MHD or CGLMHD). The notation $\hbar_M$ is adopted here purposely because it shares the same dimension with the Planck constant $\hbar$ and possess similar physical meaning.

\subsection{Conservation of Total Wave Action: Theory}
In the derivation above, a fundamental assumption is that $\mathscr{L}_A$, $\mathscr{L}_S$, $\mathscr{L}_F$ are independent with each other, which is questionable at degeneration point ($c_s=v_a, \vec k \parallel \vec B_0$). At the degeneration point, all three modes (Alfv\'en, Slow, Fast) propagate at the same phase velocity, and hence wave-wave interaction is possible. Detailed analysis shows that at the degeneration point, there are three mode-conversion mechanisms: degeneracy, linear mode conversion, resonance. The first mechanism is degeneracy of magnetosonic modes: At the degeneration point, the concept of ``Fast'' and "Slow" is ill-defined for parallel waves, and hence Fast and Slow waves would be indistinguishable from each other, i.e. an "identity crisis". Passing through the degeneration point,  the originally "Slow" wave would become "Fast" wave due to the adrupt change of the displacement polarization vector. Note that because this process happens on the $\vec k - \vec B_0$ plane, degeneracy is only possible for magnetosonic modes. The second mechanism is linear mode conversion [see e.g. \cite{swanson_theory_1998,swanson_plasma_2003, mcdougall_new_2007}]: at the degeneration point, due to the rapid change of eigenvectors, the projection of the disturbance on the each of the two magnetosonic eigenvectors change; Therefore, the initially monochromatic magnetosonic mode would be continuously linearly transformed to the mix of both slow and fast mode, until the background conditions evolve to be sufficiently distant from the degeneration point. The third mechanism is resonance: The linearly polarized Alfv\'en wave would resonate at the degeneration layer ($c_s=v_a$) to convert the wave energy into sonic modes [see e.g. \cite{hollweg_density_1971,stefani_mode_2021} and references therein], which is a candidate for chromosphere heating at the magnetic canopy [see \cite{hollweg_alfven_1982,bogdan_waves_2003}]. For all three mechanisms, the mode conversion processes are transient, and hence dissipation is negligible. Therefore, for Fast (and Slow) mode, the conversion process can be illustrated phenomenologically as:
\begin{eqnarray}
    \begin{aligned}
        \frac{E_F}{\omega_r} \xrightarrow[\mathrm{Linear\ Mode\ Conversion}]{\mathrm{Degeneracy}} \frac{E'_F}{\omega_r}+\frac{E'_S}{\omega_r}\\
        E_F = E'_F+E'_S
    \end{aligned}
    \label{degeneracy}        
\end{eqnarray}
where $E_{()}$ and $E'_{()}$ are wave energy before and after degeneration point respectively, and $\omega_r$ is intrinsic wave frequency at the degeneration point. Whereas for Alfv\'en mode:
\begin{eqnarray}
    \begin{aligned}
        \frac{E_A}{\omega_r} \xrightarrow{\mathrm{Resonance}} \frac{E'_F}{\omega_r}+\frac{E'_S}{\omega_r}+\frac{E'_A}{\omega_r}\\
        E_A = E'_F+E'_S+E'_A
    \end{aligned}
    \label{resonance}        
\end{eqnarray}
In the conversion process, the total wave energy is conservatively reallocated among corresponding degrees of freedom (eigenmodes), and hence the exchange of wave action is also conservative. Passing through degeneration point, the wave action for each degree of freedom:
\begin{eqnarray}
\hbar'_{M}=\frac{E'_{M}}{\omega_{M}}=const.
\label{h=E/w}
\end{eqnarray}
would be independently conserved. And hence the total wave action:
\begin{eqnarray}
    \hbar_{tot}=\sum_{M=A,S,F}\frac{E_{M}}{\omega_{M}}=\sum_{M=A,S,F}\frac{E'_{M}}{\omega_{M}}=const.
    \label{htot}
\end{eqnarray}
is conserved. In short, we conjecture that for MHD small-amplitude WKB perturbations, the {\it total of wave action summed over all interacting modes is a universally conserved quantity}.

\section{Simulation Results}\label{sec:result}

\subsection{Simulation Setup and Diagnostics}

\begin{figure}[h]
    \centering
    \setlength\fboxsep{0pt}
    \setlength\fboxrule{0.0pt}
    \includegraphics[width=1.0\textwidth]{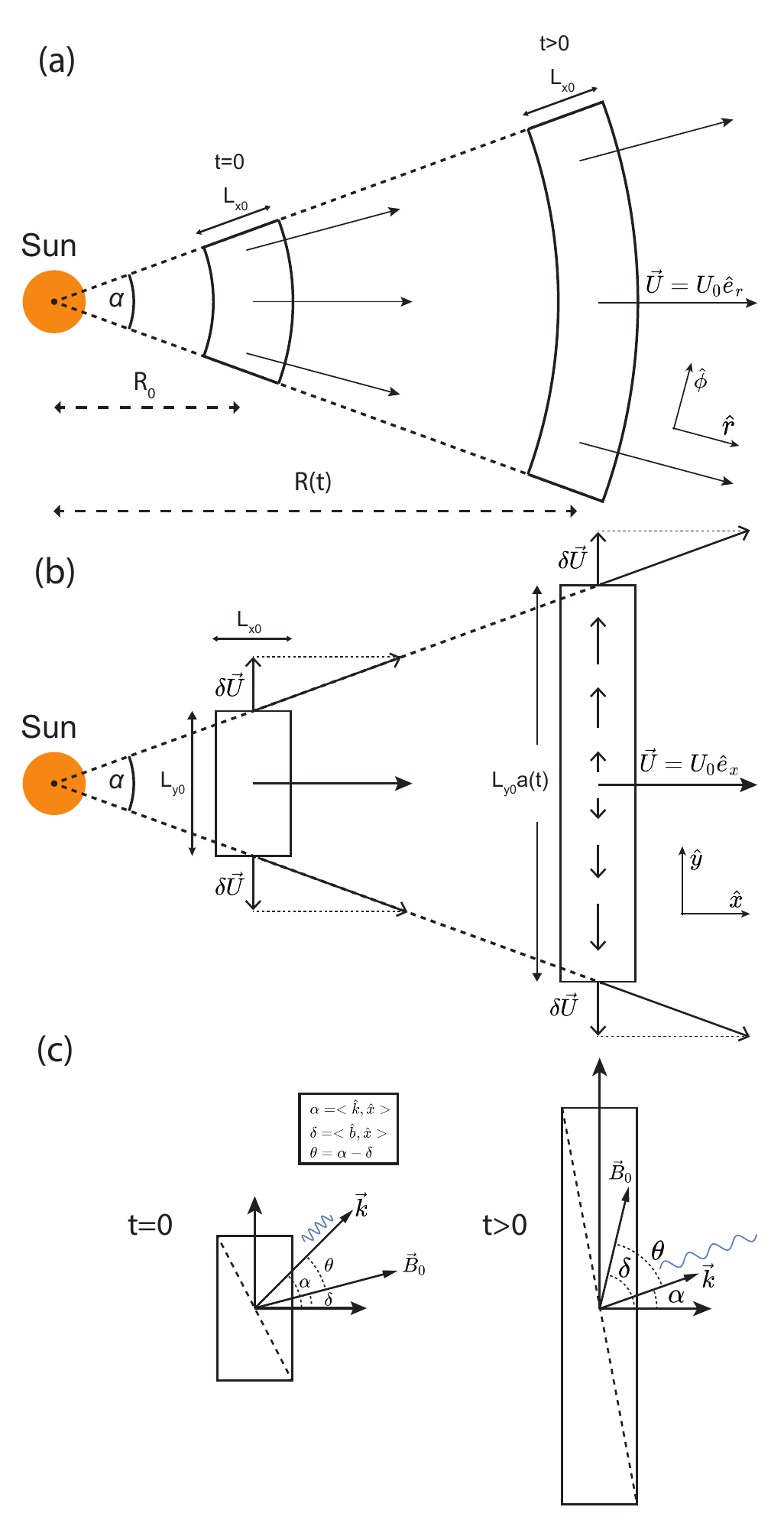}
    \caption{
        Sketch of the evolution of a plasma volume advected by a spherical wind with constant speed. (a) Exact evolution, (b) approximate evolution in the limit of small angular size (Expanding Box Model), and (c) transformation of a parallel wave ($\vec k \parallel \vec B_0$) into an oblique wave. $\vec B_0$ turns away from radial, whereas $\vec k$ turns towards radial. 
    }
    \label{fig:EBM}
\end{figure}

We conduct simulations with Expanding Box Model (EBM) formulated by [\cite{velli_mhd_1992}, \cite{grappin_nonlinear_1993, grappin_waves_1996}] and implemented by [\cite{shi_propagation_2020}]. The code is pseudo-spectral, using Fast Fourier Transform to calculate spatial derivatives and 3rd order explicit Runge-Kutta method to integrate in time. We do not add explicit viscosity or resistivity but adopt a numerical filter that adaptively dissipate shocks formed in the simulations. The simulation setup is illustrated in Figure \ref{fig:EBM}. The simulation domain is 1D with 256 grid points and comoves with the background solar wind at the speed of $U_0=400\ \mathrm{km/s}$. For each run, we initialize the simulation domain with uniform background magnetic field $\vec B_0$, pointing $\theta_0$ w.r.t. the radial direction, and run the simulation  from 0.1 AU to 1.0 AU. Velocity has unit $u^*=150 \mathrm{km/s}$, length has unit $L^*=0.012\mathrm{AU}$, and number density has unit $n^*=200 \mathrm{cm}^{-3}$, and thus magnetic field has unit of $u^*\sqrt{\mu_0 m_p n^*}=97.25\mathrm{nT}$, where $m_p$ is proton mass. The adiabatic gas constant is chosen to be $\gamma=5/3$. Different from the regular EBM, the simulation domain in our model is rotated by an initial angle $\alpha$ with respect to the radial direction, i.e., the grid points used in this study are distributed on an axis $\hat{e}_{x^\prime}$ such that the angle between $\hat{e}_r$ (the radial direction) and $\hat{e}_{x^\prime}$ is $\alpha$ initially. As the expansion effect will stretch the plasma volume in the direction perpendicular to $\hat{e}_r$, the axis $\hat{e}_{x^\prime}$ will rotate away from the radial direction, i.e. $\alpha$ will increase with time (see \cite{shi_propagation_2020} for more details).

We initialize simulations with small amplitude monochromatic Alfv\'en, Slow, and Fast wave with same wavevector $\vec k$, and vary only the initial background magnetic field modulus $|\vec B_0|$. At each time step, the wave vector is {\it a priori} determined by linear theory [\cite{volk_propagation_1973}], turning gradually towards radial:
\begin{eqnarray}
    &\vec{k}(t) = (k_{0x}, k_{0y}/a(t), 0) 
    \\
    &a(t) = \frac{R(t)}{R_0}=1+\frac{U_0}{R_0} \cdot t
\end{eqnarray}
where $a(t)$ is the expansion factor and $R_0=0.1\mathrm{AU}$. Then we extract other background quantities including $\rho_0(t)$, $p_0(t)$, and $\vec B_0(t)$ by averaging over the simulation domain. It is noteworthy that $\vec B_0(t)$, per conservation of magnetic flux, turns gradually away from radial over time (Parker Spiral):
\begin{eqnarray}
    &\vec{B_0}(t) = (B_{0x}/a(t)^2, B_{0y}/a(t), 0)
\end{eqnarray}
Given $\vec k(t)$ and other averaged background quantities, we can derive various useful quantities as diagnostics. The wave energy density is calculated by:
\begin{eqnarray}
    \mathscr{E}_w = \frac{1}{2}\langle\rho\rangle \langle(\Delta \vec u)^2\rangle
    +
    \frac{\langle(\Delta p)^2\rangle}{2 \langle\rho\rangle c_s^2}
    +
    \frac{\langle(\Delta \vec B)^2\rangle}{2\mu_0}
\end{eqnarray}
where $c_s^2=\gamma\frac{\langle p\rangle}{\langle \rho\rangle}$, $\langle()\rangle$ is the average of $()$ in the simulation domain, and $\delta () = () - \langle ()\rangle$. 

After that we need to decompose the wave energy into different degrees of freedom (Alfv\'en, Slow, and Fast mode). We first decompose the kinetic part of the wave energy density because the eigen-polarization of $\delta\vec u$ of the three eigenmodes form an orthogonal triad. And for small amplitude WKB waves, our discussion in section \ref{sec:theory} shows that $\mathscr{L}=0$, which indicates equi-partition between the kinetic ($\mathscr{E}_{k}=\frac{1}{2}\langle\rho\rangle \langle(\Delta \vec u)^2\rangle$) and potential (elastic+magnetic) ($\mathscr{E}_{p}+\mathscr{E}_{m}=\frac{\langle(\Delta p)^2\rangle}{2 \langle\rho\rangle c_s^2}+\frac{\langle(\Delta \vec B)^2\rangle}{2\mu_0}$) energy. Therefore, we can decompose the wave energy density via:
\begin{eqnarray}
    \mathscr{E}_{w,(A,S,F)}=\mathscr{E}_{k,(A,S,F)}/\mathscr{E}_k * \mathscr{E}_w
\end{eqnarray}
And with eigen-frequencies $\omega_{A,S,F}$ of each mode, we obtain the wave action for each mode:
\begin{eqnarray}
    \hbar_{A,S,F}=\int_{V(t)} \frac{\mathscr{E}_{w,(A,S,F)}}{\omega_{A,S,F}}d\nu=\frac{E_{w,(A,S,F)}}{\omega_{A,S,F}}
    \label{Ew,A,S,F}
\end{eqnarray}
where $V(t)$ is the volume of the "Expanding" simulation domain at time $t$ and $E_{\omega,(A,S,F)}$ is the integrated wave energy enclosed by the simulation domain. Finally, we have the total wave action:
\begin{eqnarray}
    \hbar_{tot} = \hbar_A + \hbar_S + \hbar_F
    \label{eqn:hbar tot}
\end{eqnarray}
The conservation of total wave action states that: $\hbar_{tot}=const.$, and thus we diagnose each run with the normalized total wave action $\tilde{\hbar}_{tot}(t)$:
\begin{eqnarray}
    \begin{aligned}
        \tilde{\hbar}_{tot}(t)
        &=\hbar_{tot}(t)/\hbar_{tot}(0)\\
        &=\tilde{\hbar}_{A}(t)
        +\tilde{\hbar}_{S}(t)
        +\tilde{\hbar}_{F}(t)
        \label{eqn:normalized hbar tot}
    \end{aligned}
\end{eqnarray}
This is the primary diagnostic for our simulations.

\subsection{Conservation of Total Wave Action: Simulation}
\begin{figure}[h]
    \centering
    \includegraphics[width=1.0\textwidth]{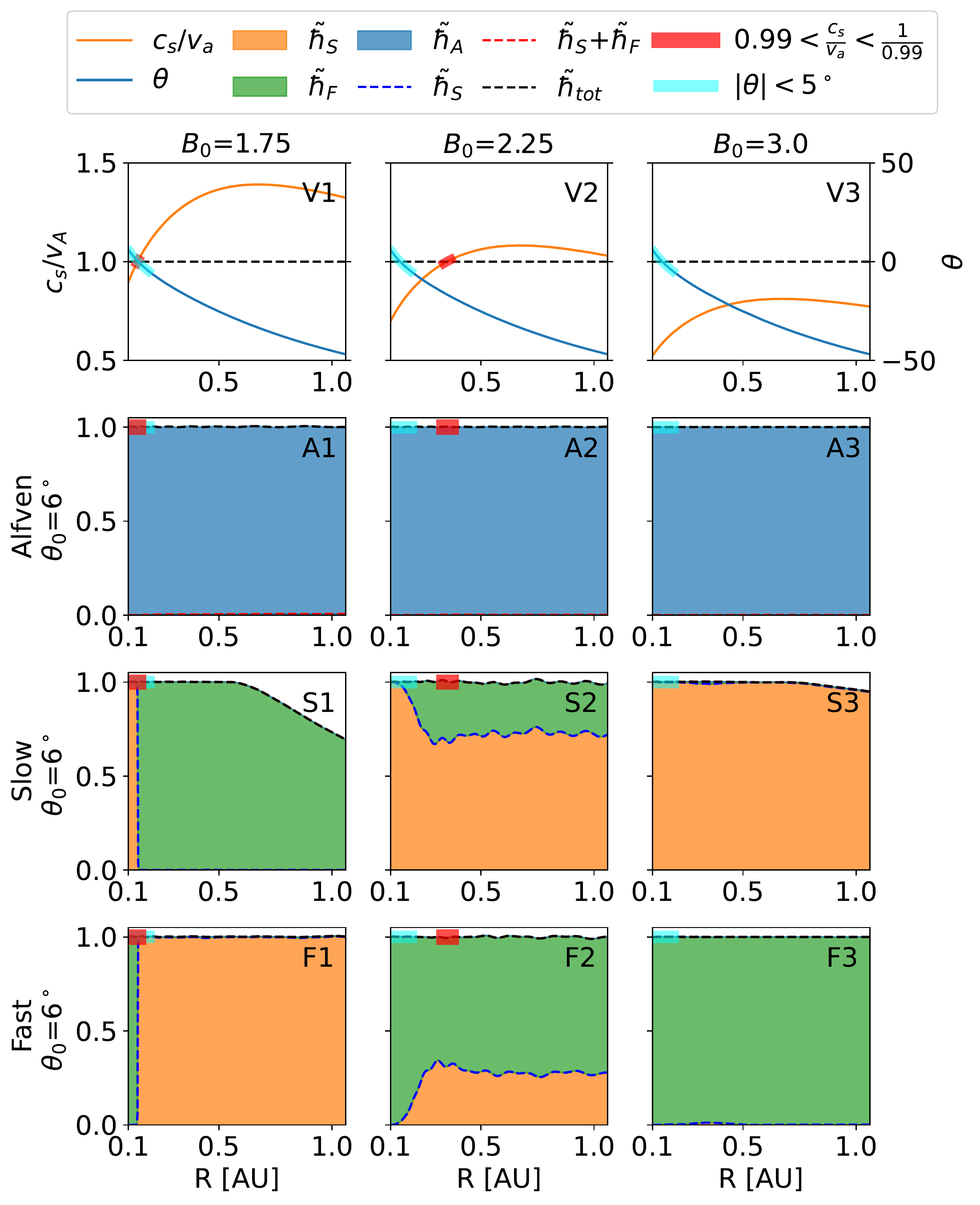} 
    \caption{
        The evolution of normalized total wave action with initial monochromatic Alfv\'en (A1-A3), Slow (S1-S3), and Fast wave (F1-F3) in expanding box simulation, together with resonance/degeneracy condition (V1-V3). All runs are initialized with $\delta_0=<\vec B_0, \hat r>|_{t=0}=6^\circ$, $\alpha_0=<\vec k, \hat r>|_{t=0}=12^\circ$, and $\theta_0=\alpha_0-\delta_0$, varying only $|\vec B_0|$. The evolution of plasma parameters $c_s/v_a$ and $\theta=<\vec B_0, \vec k>$ are plotted in the top row with orange and blue lines, and the region close to resonance are highlighted with overlaid red and cyan bar on all panels. Rows 2-4 show the radial evolution of normalized wave actions with different colors, respectively initialized with monochromatic Alfv\'en, Slow, and Fast wave. The color in the panels indicate the normalized wave action for Alfv\'en/Slow/Fast mode denoted with $\tilde{\hbar}_{A}$/$\tilde{\hbar}_{S}$/$\tilde{\hbar}_{F}$, and they are stacked together, as indicated by dashed lines ($\tilde{\hbar}_{S}$,$\tilde{\hbar}_{S}+\tilde{\hbar}_{F}$), and finally into the normalized total wave action $\tilde{\hbar}_{tot}$.
    }
    \label{fig:hbar map}
\end{figure}

To prove our conjecture on conservation of total wave action, the initial conditions are carefully selected so that the resonance conditions can be satisfied perfectly or partially in the simulation. Figure \ref{fig:hbar map} shows nine simulation runs of monochromatic Alfv\'en, Slow, and Fast waves with three different initial $|\vec B_0|$ (hence Alfv\'en speed $v_a$). All runs are initialized with uniform $\vec B_0$ with $\delta_0=<\vec B_0, \hat r>|_{t=0}=6^\circ$, and initial wave vector $\vec k$ with $\alpha_0=<\vec k, \hat r>|_{t=0}=12^\circ$, both pointing counterclockwise w.r.t. radial $\hat r$ ($\alpha=<\vec k, \hat r>, \delta=<\vec B_0, \hat r>, \theta=<\vec k, \vec B_0>$, also see Figure \ref{fig:EBM}). To understand the evolution of monochromatic waves, we show in each panel of Figure \ref{fig:hbar map} the normalized total wave action $\tilde{\hbar}_{tot}$ defined in (\ref{eqn:normalized hbar tot}) and its composition in three different colours: $\tilde{\hbar}_A$ (Alfv\'en, Blue), $\tilde{\hbar}_S$ (Slow, Orange), $\tilde{\hbar}_F$ (Fast, Green). The resonance criteria, $c_s/v_a$ and $\theta=<\vec k, \vec B_0>$ are shown in the top row, and resonant windows are highlighted with red and cyan bars, also overlaid in all panels to indicate the same periods. 

As shown in Figure \ref{fig:EBM}(c), $\vec B_0$ turns gradually away from radial, whereas $\vec k$ turns gradually towards radial over time, and thus with our setup ($\alpha_0>\delta_0$), two vectors will coincide as the wave propagating outwards. The three different initial $|\vec B_0|$ are carefully selected to represent perfect degeneration point passing ($c_s=v_a, \vec k \parallel \vec B_0$ are perfectly satisfied simultaneously), partial degeneration point passing (Both $c_s=v_a, \vec k \parallel \vec B_0$ are satisfied, but not simultaneously), and miss (one of the resonant criteria is not satisfied), shown respectively in column 1-3 in Figure \ref{fig:hbar map}.

Results show that all runs start with conserved wave action (only one color is presented at a given time step, vertical intersection), and some of the runs (S1, S2, F1, F2) subsequently convert to other modes. Specifically, run S1 passes through the degeneration point perfectly (overlapping red and cyan overhead bars) at around 0.1 AU and hence converts completely from Slow mode (orange) to Fast mode (green), and vice versa for run F1. On the other hand, run S2 passes through the degeneration point semi-perfectly, and thus run S2 converts partially from Slow mode to Fast mode, and vice versa for run F2. Most importantly, all of the four runs, albeit having mode conversion, maintain an almost constant total wave action all over the evolution. Especially for run S2 and F2, after the transient mode conversion phase, the slow mode and fast mode part of the wave coexist, and the wave action for both modes are independently conserved.

Other runs (A1-A3, S3, F3) present no sign of mode conversion and therefore maintain a constant total (albeit monochromatic) wave action. One may notice that for runs S1 and S3, the total wave action decreases significantly towards the end (R $>$ 0.5 AU). This is due to dissipation of shock formed via wave steepening. 

\section{Discussion}\label{sec:discussion}

In this section we give a short discussion on the stability of Alfv\'en wave and the mechanisms of mode-conversion seen in the magnetosonic modes.
\subsection{Stability of Alfv\'en Wave}
As shown in Figure \ref{fig:hbar map}, Alfv\'en wave appears to be more stable than magnetosonic waves. A simple explanation to this is that Alfv\'en wave is a transverse wave and hence per Burgers' equation, Alfv\'en wave does not resonates with itself. More specifically, the inviscid Burgers' equation is written as:
\begin{eqnarray}
    \frac{\partial \vec u}{\partial t}+\vec u \cdot \nabla \vec u=0
    \label{eqn:burgers equation}
\end{eqnarray}
For Alfv\'en mode, as a transverse wave, the convective term is zero:
\begin{eqnarray}
    \vec u \cdot \nabla \vec u = 0
\end{eqnarray}
Hence no self-resonance is present for Alfv\'en wave. Moreover, the displacement vector of Alfv\'en wave is perpendicular to the $\vec k-\vec B_0$ plane. It is hence extremely hard for Alfv\'en wave to convert to the two magnetosonic modes with linear mode conversion. Therefore, the only viable mechanism in our setup for Alfv\'en wave to convert to other magnetosonic modes is through Alfv\'en resonance [\cite{hollweg_density_1971,stefani_mode_2021}]. The effectiveness of the resonance is proportional to both wave amplitude and interaction time. It is hence very hard for Alfv\'en wave to exhibit observable nonlinear effect if the wave amplitude is small and is propagating in expanding medium. On the other hand, if we abandon the expansion effects and run the simulation without expansion effect, or increase the wave amplitude, we may achieve significant mode conversion for the Alfv\'en wave. Therefore, it is interesting to see whether the total wave action is a better-conserved quantity than single-mode wave action with the presence of significant mode-conversion for Alfv\'en wave.

\begin{figure}[h]
    \centering
    \includegraphics[width=1.0\textwidth]{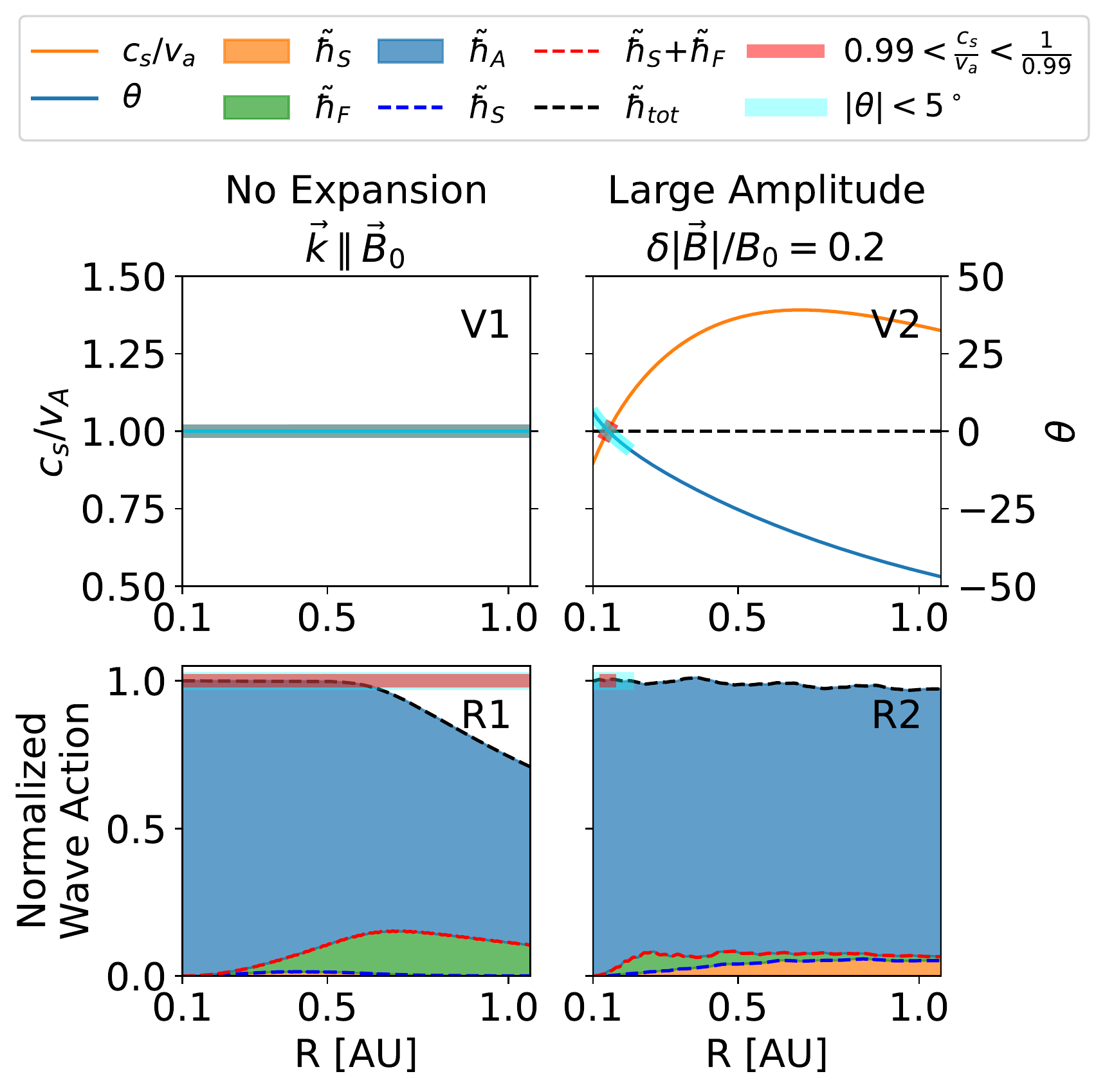} 
    \caption{
        The evolution of normalized total wave action with initial monochromatic Alfv\'en waves. Run R1: small amplitude, always resonant, and has no expansion; Run R2: large amplitude, transient perfect resonant, and has expansion. Resonant conditions are shown in V1/V2. All legends are identical to Figure \ref{fig:hbar map}. The total wave action is conserved for both runs.
    }
    \label{fig:alfven resonance}
\end{figure}

Figure \ref{fig:alfven resonance} demonstrates two simulation runs, showing respectively small-amplitude Alfv\'en wave without expansion effect (R1), and large-amplitude Alfv\'en wave with expansion effect (R2). Simulation results show that both abandoning expansion effect and increasing amplitude can induce significant mode-conversion (resonance). Moreover, the normalized total wave action plots (R1, R2) clearly show that, albeit with significant resonance, the total wave action remains almost constant until shock dissipation intensify.

\subsection{Magnetosonic Wave Mode Resonance}
The mode conversion processes of magnetosonic waves in Figure \ref{fig:hbar map}, panel S2/F2 are significantly different from the complete mode conversion in panel S1/F1. In fact, they exemplify two distinct mode conversion mechanisms, i.e. degeneracy and linear mode conversion [see e.g. \cite{zhugzhda_magnetogravity_1979,zhugzhda_conversion_1981,zhugzhda_transformation_1982,zhugzhda_linear_1982,cairns_unified_1983,swanson_theory_1998,cally_note_2001,swanson_plasma_2003,mcdougall_new_2007,mcdougall_mhd_2007} and references therein]. Degeneracy happens only when degeneration point ($c_s=v_a, \vec k \parallel \vec B_0$) passing is perfect, and hence is very rare. Linear mode conversion happens within a small region around the degeneration point, where the dispersion relation of Slow and Fast mode coincides, and hence is more universal.

\begin{figure}[h]
    \centering
    \includegraphics[width=1.0\textwidth]{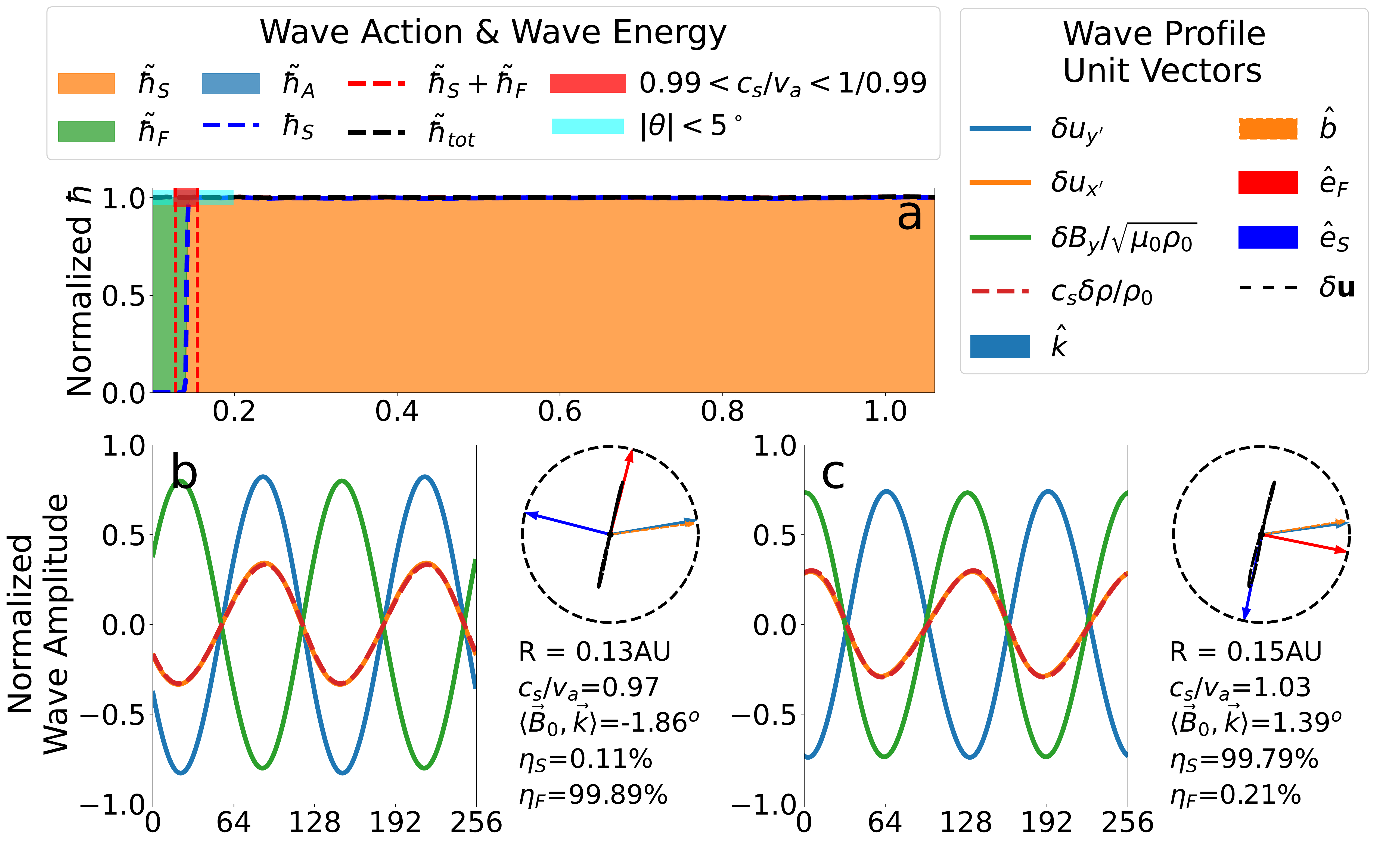} 
    \caption{
        The detailed evolution of run F1 with wave profile is shown here. Panel a is identical to panel F1 in Figure \ref{fig:hbar map}; x axis is radial distance to sun (R [AU]). Panel b and c are the wave profile at two time steps indicated by the two red, vertical dashed line in panel a; x axis is grid points. The legends of panel a are identical to legends in Figure \ref{fig:hbar map}. The blue, orange solid line in panel b and c are flow speed fluctuation (displacement) amplitude along x' (parallel to $\vec k$) and y' (coplanar with $\vec k$ and $\vec B_0$) direction; green solid line is the normalized magnetic fluctuation amplitude; and the red dashed line is the density fluctuation amplitude. In the radar-like arrow plot, all arrows are unit vectors: the light blue arrow is the wave vector $\vec k$; the orange dashed arrow is the background magnetic field $\vec B_0$; the red and deep blue arrows are unit vectors of displacement of Fast and Slow mode respectively. The black dashed closed loop is the trajectory of displacement from edge to edge in the simulation domain (trajectory of the blue and orange line in corresponding wave profile panel on the left). The texts in panel b and c are important information of the time frame, where $\eta_{S/F}=\epsilon_{w,S/F}/(\epsilon_{w,S}+\epsilon_{w,F})$ is the ratio of the wave energy belongs to either Slow or Fast mode. For example, in panel b, the black dashed closed loop is parallel to the red arrow, indicating that the wave is pure fast mode; whereas in panel c, the loop is parallel to the blue arrow, indicating that the wave is pure slow mode. By checking both the radar plot and the value of $\eta_{S/F}$, from panel b to c, we clearly witness a mode degeneracy of magnetosonic modes.
    }
    \label{fig:complete mode conversion run F1}
\end{figure}

The complete conversion in panels S1/F1 can be simply explained by the sudden change of the displacement polarization upon passing through the degeneration point, i.e. degeneracy of wave modes. The detailed evolution of run F1 is shown in Figure \ref{fig:complete mode conversion run F1}. Two wave profiles at two time steps adjacent to the mode conversion point are shown for comparison. Before entering the degeneration point, the displacement vector's trajectory (Lissajous curve) from edge to edge in the simulation domain (dark dashed close loop, radar plot, panel b) is parallel to the fast mode displacement polarization (red arrow); and in the meantime the wave vector (blue arrow) and the background magnetic field (orange dashed arrow) are very closely aligned with each other. Passing through the degeneration point (see the slight change of $<\vec B_0, \vec k>$ before and after the degeneration point), the wave profile is hardly modified (panel b/c blue, orange, green, and red dashed line), but the polarization vectors have an abrupt change (sudden change of red/blue vectors in panel b/c, radar plot) because the meaning of "Fast" and "Slow" switches at degeneration point, and hence the projection of the displacement vector's trajectory (dark dashed line, radar plot) on the two polarization vector (red/blue vectors, radar plot) has an abrupt change.

\begin{figure}[h]
    \centering
    \includegraphics[width=1.0\textwidth]{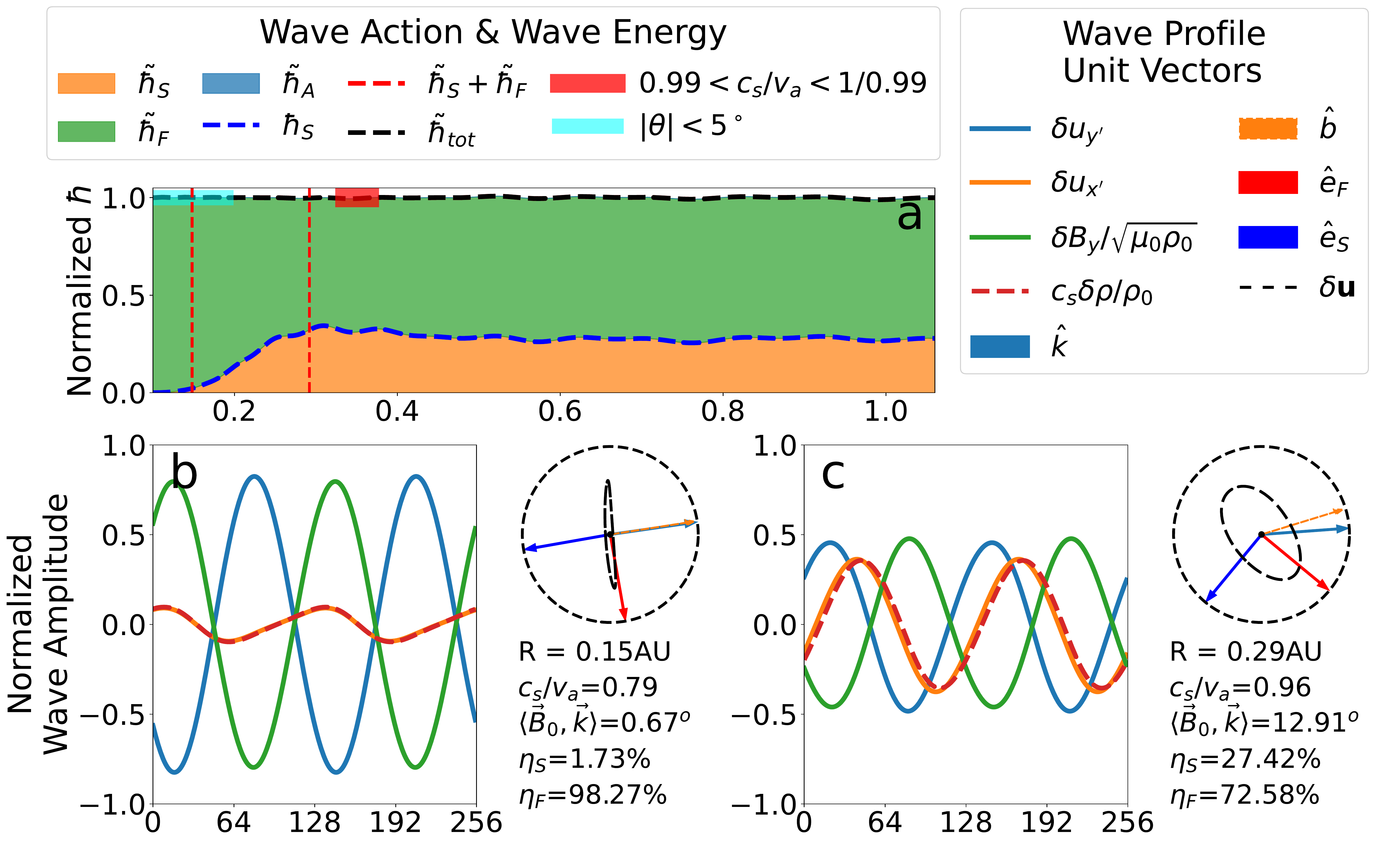} 
    \caption{
        The detailed evolution of run F2 with wave profile shown. Panel a is identical to panel F2 in Figure \ref{fig:hbar map}. All legends are identical to Figure \ref{fig:complete mode conversion run F1}. Note that the black dashed loop in panel b and c are trajectory of displacement vector from edge to edge in the simulation domain. In panel b, the loop is mostly parallel to the red arrow, indicating that the wave is mostly pure fast wave (also see the time step indicated by the first red, vertical dashed line in panel a, is almost all green); in panel c, the loop has projection on both red and blue vector, indicating that the wave is a mixed slow and fast wave (also see the time step indicated by the second red, vertical dashed line in panel a, is mixed green and orange). Moreover, by comparing the value of $\eta_{S/F}$ in panel b and c, we obviously witness a magnetosonic linear mode conversion from panel b to c.
    }
    \label{fig:partial mode conversion run F2}
\end{figure}

For comparison, the detailed evolution of run F2 is shown in Figure \ref{fig:partial mode conversion run F2}. As we can see in panel b and c, the linearly polarized Fast wave started to convert to slow mode via linear mode conversion (see panel a in Figure \ref{fig:partial mode conversion run F2}, the growing ratio of orange area (slow mode) from 0.1 AU and 0.3 AU). Such linear mode conversion happens because around the degeneration point, the eigen-vectors of magnetosonic modes are changing rapidly, and therefore the system becomes non-WKB. The rapid change of the eigen-vectors changes the mixing ratio of slow and fast mode (see the radar plots in panel b and c, depicting the wave profiles at two time steps indicated by two red vertical dashed line in panel a). Subsequently, because of the phase speed difference between two modes, the Lissajous curve of the wave change from an linearly polarized wave (thin dashed black close loop in radar plot, panel b) to a circularly polarized wave (oval-like dashed black close loop in radar plot, panel c). Note that the oval-like Lissajous curve indicates that the two wave modes have similar frequencies, further confirming the mode conversion process is linear (or else would transport wave energy to higher wave number).

\section{Summary}\label{sec:summary}

Half a century ago, the theory of wave action conservation is devised to describe the nonlinear evolution of WKB waves [see \cite{whang_alfven_1973}, \cite{whitham_general_1965}, \cite{f_p_bretherton_wavetrains_1968}, \cite{dewar_interaction_1970}]. However, the classical theory fails to predict the mode-conversion happening close to the MHD degeneration point ($c_s=v_a, \vec k \parallel \vec B_0$). In this paper, we have shown that although mode conversion violates the conservation of wave action for infinitely long monochromatic MHD wave trains propagating in the expanding solar wind, the total of wave action summed over all interacting modes (Alfv\'en, Slow and Fast) remains a universally conserved quantity. 1D MHD simulation with the Expanding Box Model (EBM) [\cite{velli_mhd_1992}, \cite{grappin_nonlinear_1993}, \cite{grappin_waves_1996}, \cite{shi_propagation_2020}] demonstrate this and further reveal that there are three distinct mode conversion mechanisms: degeneracy, linear mode conversion and resonance. A simple physical picture is that, due to the expansion of the medium, wave vector $\vec k$ turns towards radial, and background magnetic field $\vec B_0$ turns away from radial per Parker Spiral. Hence with special setup, when the two vectors align with each other and in the mean time sound speed $c_s$ and Alfv\'en speed $v_a$ becomes nearly identical, all three mode conversion mechanisms become possible. 

Degeneracy is due to the fact that the concepts of "Fast" and "Slow" become ill-defined at the degeneration point for parallel waves, and hence passing through the degeneration point, the originally "Slow" wave can become "Fast" due to the abrupt change of the displacement polarization vector (see Figure \ref{fig:complete mode conversion run F1}). Therefore, degeneracy can only happen for magnetosonic modes, and is not applicable to Alfv\'en mode. Linear mode conversion on the other hand is more universal for magnetosonic waves [see e.g. \cite{zhugzhda_magnetogravity_1979, zhugzhda_conversion_1981, zhugzhda_transformation_1982,zhugzhda_linear_1982,cally_note_2001, mcdougall_mhd_2007,mcdougall_new_2007,mcdougall_mhd_2009} for similar linear mode conversion for magnetogravity waves at the magnetic canopy in solar chromosphere]. Finally, resonance can happen for Alfv\'en mode, where the well-known Alfv\'en resonance can generate secondary Slow and Fast waves [see \cite{hollweg_density_1971} or Appendix-A, and simulation in Figure \ref{fig:alfven resonance}]. In short, the mode conversion process and the conservation of total wave action can be summarized as:
\begin{eqnarray}
    \frac{E_M}{\omega_r}\xrightarrow[\mathrm{Resonance}]{\mathrm{Degeneracy}}\sum_M \frac{E'_M}{\omega_r}
\end{eqnarray}
where $E_M$ and $E'_M$ are wave energy before and after resonance/degeneracy of mode M, and $\omega_r$ is resonance frequency. 

We believe our proposed physical model is generally applicable to any fluid system because: (a) wave action is a universal concept, regardless of system description; (b) our mathematical description on conservation of total wave action is general, without concerning the details of MHD; (c) All three mode-conversion mechanisms are universal phenomena regardless of fluid description. Hence by providing simple, intuitive physical picture for mode conversion, our model generalizes the classical theory of wave action conservation.

\section*{Appendix-A}\label{sec:appendix-A}

Here for completeness, we give a short derivation on the variation principle for wave action. For a slowly varying (WKB) wavetrain, the dominant local amplitude, frequency, and wavenumber are governed by the variational principle:
\begin{equation*}
    \delta \int \mathscr{L}(\tilde a, -\theta_t, \theta_x) dx dt = 0
\end{equation*}
subject to infinitesimal variation $\delta \theta(x,t)$ which vanish at infinity. Variation with respect to $\theta$ yields:
\begin{equation*}
    \begin{aligned}
        \int &\left[
            \frac{\partial \mathscr{L}}{\partial \theta_t}\delta\left(\frac{\partial \theta}{\partial t}\right)
            +
            \frac{\partial \mathscr{L}}{\partial \theta_x}\delta\left(\frac{\partial \theta}{\partial x}\right)
        \right]dx dt\\
        &=\int\left[
            \frac{\partial \mathscr{L}}{\partial \theta_t}\left(\frac{\partial \delta \theta}{\partial t}\right)
            +
            \frac{\partial \mathscr{L}}{\partial \theta_x}\left(\frac{\partial \delta \theta}{\partial x}\right)
        \right]dx dt\\
        &=\int \left[
            \frac{\partial }{\partial t}\left(
                \frac{\partial \mathscr{L}}{\partial \theta_t} \delta \theta
            \right)
            +
            \frac{\partial }{\partial x}\left(
                \frac{\partial \mathscr{L}}{\partial \theta_x} \delta \theta
            \right)\right.\\
            &\qquad\left.-
            \frac{\partial }{\partial t}\left(
                \frac{\partial \mathscr{L}}{\partial \theta_t}
            \right)\delta \theta
            -
            \frac{\partial }{\partial x}\left(
                \frac{\partial \mathscr{L}}{\partial \theta_x}
            \right)\delta \theta
        \right]dx dt\\
        &=
        \underbrace{
        \int\left.
                \frac{\partial \mathscr{L}}{\partial \theta_t} \delta \theta
            \right|_t dx}_{0}
        +
        \underbrace{
        \int\left.
                \frac{\partial \mathscr{L}}{\partial \theta_x} \delta \theta
            \right|_x dt}_{0}\\
        &\qquad+
        \int\left[
            \frac{\partial }{\partial t}\left(
                \frac{\partial \mathscr{L}}{\partial \omega}
            \right)
            -
            \frac{\partial }{\partial x}\left(
                \frac{\partial \mathscr{L}}{\partial k}
            \right)
        \right]\delta \theta dx dt
    \end{aligned}
\end{equation*}
Hence finally we obtain:
\begin{equation*}
    \frac{\partial }{\partial t}\left(
    \frac{\partial \mathscr{L}}{\partial \omega}
    \right)
    -
    \frac{\partial }{\partial x}\left(
    \frac{\partial \mathscr{L}}{\partial k}
    \right)=0
\end{equation*}

\section*{Appendix-B}\label{sec:appendix-B}

Following equation (13) in [\cite{hollweg_density_1971}]
, for a monochromatic linearly polarized Alfv\'en wave propagating parallel to $\vec B_0$, the secondary density fluctuation is driven by the non-uniform magnetic pressure:
\begin{eqnarray}
    \frac{\partial^2 \delta \rho}{\partial t^2}-c_s^2 \frac{\partial^2 \delta \rho}{\partial x^2}=\frac{\partial^2}{\partial x^2}\left(
        \frac{\delta B_z^2}{2\mu_0}
    \right)
    \label{eqn:alfven resonance equation}
\end{eqnarray}
where $\delta B_z(x,t)=\tilde{B}_z \cos[k (x- v_a t)]$, and $\delta \rho$ is density fluctuation induced by first order Alfv\'en wave. When $c_s\ne v_a$ the usual particular solution to this equation is:
\begin{eqnarray}
    \delta \rho_p(x,t)=-\frac{\tilde{B}_z^2 }{4 \mu_0(c_s^2-v_a^2)}\cdot \cos[2k(x- v_a t)]
    \label{eqn:normal particular solution}
\end{eqnarray}
However, $c_s=v_a$ is a degeneration point and in this case, equation (\ref{eqn:normal particular solution}) has the particular solution:
\begin{eqnarray}
    \delta \rho_{p,r}(x,t)=\frac{\tilde{B}_z^2 k }{4 \mu_0 v_a}\cdot t\sin[2 k (x-v_a t)]
\end{eqnarray}
which grows linearly in time. The resonance strength is proportional to interaction time (time satisfying the resonance condition) and wave amplitude.

\bibliography{sample631}{}
\bibliographystyle{aasjournal}



\end{CJK*}
\end{document}